\documentclass[a4paper,11pt]{article}
\usepackage{pos}
\newcommand{\Bd}{B_d^0}
\newcommand{\Bs}{B_s^0}
\newcommand{\barBs}{\bar B_s^0}
\newcommand{\KS}{K_{\rm S}}
\newcommand{\BR}{\mathcal B}
\newcommand{\Adir}{\mathcal A_{\rm CP}^{\rm dir}}
\newcommand{\Amix}{\mathcal A_{\rm CP}^{\rm mix}}
\newcommand{\ADG}{\mathcal A_{\rm CP}^{\Delta\Gamma}}
\newcommand{\eps}{\epsilon}
\newcommand{\SM}{\mathrm{SM}}
\newcommand{\NP}{\mathrm{NP}}
\newcommand{\NF}{\mathrm{NF}}
\newcommand{\Uobs}{\mathcal U}
\newcommand{\Wobs}{\mathcal W}
\newcommand{\Zobs}{\mathcal Z}
\newcommand{\Sobs}{\mathcal S}
\newcommand{\Dobs}{\mathcal D}

\title{CP violation in $B_{(s)}\to\phi K$ decays:\\
Standard Model and isospin-dependent New Physics}
\ShortTitle{CP violation in $B_{(s)}\to\phi K$ decays}

\author[a,b]{Robert Fleischer}
\author*[a,c]{Jelle Groot}
\author[a,d]{K. Keri Vos}

\affiliation[a]{Nikhef, Science Park 105, NL-1098 XG Amsterdam, Netherlands}
\affiliation[b]{Department of Physics and Astronomy, Vrije Universiteit Amsterdam,\\
NL-1081 HV Amsterdam, Netherlands}
\affiliation[c]{Institute for Theoretical Physics Amsterdam and Delta Institute for Theoretical Physics,\\
University of Amsterdam, Science Park 904, 1098 XH Amsterdam, Netherlands}
\affiliation[d]{Gravitational Waves and Fundamental Physics (GWFP), Maastricht University,\\
Duboisdomein 30, NL-6229 GT Maastricht, Netherlands}

\emailAdd{r.fleischer@nikhef.nl}
\emailAdd{j.groot@nikhef.nl}
\emailAdd{k.vos@nikhef.nl}

\abstract{The penguin-dominated $B\to\phi K$ decays provide sensitive probes of physics beyond the Standard Model. Interpreting them at increasing experimental precision requires control of doubly Cabibbo-suppressed hadronic contributions. Using a factorization approach, we obtain Standard Model benchmarks for the branching ratios and CP-violating observables, including the hadronic phase shift in $B_d^0\to\phi K_{\rm S}$. The $B_s^0\to\phi K_{\rm S}$ channel, in which the corresponding penguin effects are not Cabibbo-suppressed, offers additional insight into these contributions. Comparing the neutral and charged $B\to\phi K$ modes further allows the construction of isospin observables. We discuss their Standard-Model expectations and sensitivity to New Physics in the $I=0$ and $I=1$ sectors. The available data are consistent with the Standard Model, while leaving substantial room for New Physics contributions.}

\FullConference{14th Edition of the Large Hadron Collider Physics Conference (LHCP2026)\\
18--22 May 2026\\
Paris, France}

\makeatletter
\gdef\PoScopyright@box{\parbox[b]{.80\textwidth}{\tiny
$\copyright$ Copyright owned by the author(s) under the terms of the Creative Commons\\
Attribution-NonCommercial-NoDerivatives 4.0 International License (CC BY-NC-ND 4.0).\\
All rights for text and data mining, AI training, and similar technologies for commercial purposes, are reserved.\\
ISSN 1824-8039. Published by SISSA Medialab.}}
\makeatother
\hypersetup{pdftitle={CP violation in B(s) -> phi K decays: Standard Model and isospin-dependent New Physics},pdfauthor={Jelle Groot},pdfsubject={LHCP2026 parallel-session proceedings}}

\begin{document}
\maketitle

\section{\texorpdfstring{\boldmath Standard Model benchmarks for $B\to\phi K$ decays}{Standard Model benchmarks for B to phi K decays}}
\label{sec:sm}

The decays $\Bd\to\phi\KS$ and $B^+\to\phi K^+$ originate from $\bar b\to \bar s s \bar s$ transitions and are dominated by penguin topologies in the Standard Model (SM). Their loop suppression makes them particularly sensitive to New Physics (NP)~\cite{Fleischer:1996penguins, Grossman:1997phiK, Grossman:2003SU3, Fleischer:2001isospin, Biswas:2024phiK}. However, interpreting increasingly precise measurements of CP violation requires accounting for hadronic contributions that are often neglected. This contribution summarises work presented in Ref.~\cite{Fleischer:2026phiK}.

Using CKM unitarity, the $B\to\phi K$ amplitudes can be parametrized as
\begin{equation}
\label{eq:amplitude-sm}
A(B\to\phi K)=\frac{\mathcal A'_P}{\sqrt{\eps}}\left[1+\eps b e^{i\theta}e^{i\gamma}\right]\ ,\qquad b e^{i\theta}=R_b\frac{P^{(ut)\prime}}{P^{(ct)\prime}}\ ,
\end{equation}
where $P^{(qt)\prime}=P'_q-P'_t$ denotes a difference of penguin amplitudes, $R_b$ is a side of the unitarity triangle, and $\gamma$ and $\theta$ are CP-violating weak and CP-conserving strong phases, respectively. The factor $\eps=\lambda^2/(1-\lambda^2)\simeq0.05$, with $\lambda=|V_{us}|$, renders the second term doubly Cabibbo-suppressed.

We estimate the hadronic matrix elements by factorizing them into matrix elements of the corresponding quark currents. The leading amplitudes then involve the $\phi$ decay constant, a $B\to K$ form factor, and non-factorizable normalizations $a_{\NF}^{0,+}=1+\delta_{\NF}^{0,+}$, where the superscripts $0$ and $+$ refer to the neutral and charged modes, respectively. Using NLO Wilson coefficients and one-loop penguin contractions of the current--current operators gives $b e^{i\theta}\simeq R_b \mathcal C_{u,\SM}/\mathcal C_{c,\SM}$, where $\mathcal C_{q,\SM}$ combines Wilson coefficients and loop functions~\cite{Fleischer:1994}. Varying the virtual gluon or photon momentum over $1/4<k^2/m_b^2<1/2$ yields
\begin{equation}
\label{eq:penguin-input}
b_{\rm incl}=0.38\pm0.04\ ,\qquad b_{\rm excl}=0.34\pm0.04\ ,\qquad \theta=(23\pm6)^\circ\ .
\end{equation}
The labels refer to inclusive and exclusive CKM determinations, while the strong phase $\theta$ is essentially insensitive to this choice. \!Unless stated otherwise, we use inclusive CKM inputs throughout this contribution, and the corresponding exclusive SM benchmark inputs are given in Ref.~\cite{Fleischer:2026phiK}.

The CP asymmetries are defined with $\Gamma(B)-\Gamma(\bar B)$ in the numerator and $\Amix$ multiplying the $+\sin(\Delta M_d t)$ term. For $\Bd\to\phi\KS$, we can define an effective mixing phase $\Delta\phi_d^{\phi\KS}$ that satisfies
\begin{equation}
\label{eq:phase}
-\Amix/{\sqrt{1-(\Adir)^2}}=\sin\!\left(\phi_d+\Delta\phi_d^{\phi\KS}\right)\ ,\qquad \Delta\phi_d^{\phi\KS}=2\eps b\cos\theta\sin\gamma+\mathcal O(\eps^2)
\end{equation}
in the absence of NP in the decay amplitude. Using $\gamma=(64.9\pm4.5)^\circ$~\cite{LHCb:2021gamma} and the mixing phase $\phi_d=(44.4^{+1.6}_{-1.5})^\circ$, extracted from $\Bd\to J/\psi\KS$ including doubly Cabibbo-suppressed penguin contributions~\cite{Barel:2023penguins}, gives the SM benchmarks
\begin{equation}
\label{eq:bd-results}
\Adir=-0.014\pm0.005\ ,\qquad \Amix=-0.723\pm0.019\ ,\qquad \Delta\phi_d^{\phi\KS}=(1.9\pm0.2)^\circ\ .
\end{equation}
The overall normalization $a_{\NF}^{0}$ cancels in these observables, but the dependence on $b$ and $\theta$ remains. The phase shift, correlated with $\Adir$ in Fig.~\ref{fig:sm-correlations} (left), becomes relevant as the experimental precision approaches a few degrees. Neglecting possible non-factorizable effects $(a_{\rm NF}^{0,+}=1)$ yields
\begin{equation}
\label{eq:br-results}
\BR(\Bd\to\phi\KS)=(4.9\pm1.1)\times10^{-6}\ ,\qquad \BR(B^+\to\phi K^+)=(10.5\pm2.4)\times10^{-6}\ .
\end{equation}
Within the SM, comparison with the measured branching ratios gives~\cite{Fleischer:2026phiK}
\begin{equation}
\label{eq:delta-results}
\delta_{\NF}^{0}=-0.14\pm0.12\ ,\qquad \delta_{\NF}^{+}=-0.09\pm0.12\ .
\end{equation}
These results show no significant deviation from the factorization framework, although non-factorizable corrections of order $10\%$ remain allowed.

\clearpage
\section{\texorpdfstring{\boldmath The $B_s^0\to\phi K_{\rm S}$ channel: amplified penguin effects}{The Bs to phi KS channel: amplified penguin effects}}
\label{sec:bs}

The $\Bs\to\phi\KS$ decay provides a complementary probe of the hadronic penguin parameters. It proceeds through a $\bar b\to\bar d s \bar s$ transition, giving the amplitude
\begin{equation}
\label{eq:bs-amplitude}
A(\Bs\to\phi\KS)=-\widetilde{\mathcal A}^{(ct)}_P\left[1-\widetilde b e^{i\widetilde\theta}e^{i\gamma}\right]\ .
\end{equation}
Measurements of the direct and mixing-induced CP asymmetries would therefore determine $\widetilde b$ and $\widetilde\theta$, using $\gamma$ and the $\barBs$ mixing phase as inputs. Unlike in Eq.~\eqref{eq:amplitude-sm}, the second term is not suppressed by $\eps$, amplifying the hadronic penguin effects relative to $\Bd\to\phi\KS$ and $B^+\to\phi K^+$. 

\begin{figure}[htb]
\centering
\includegraphics[width=.49\textwidth]{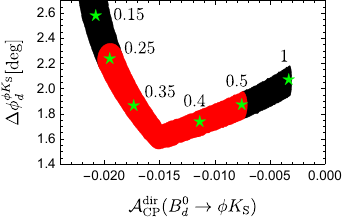}\hfill
\includegraphics[width=.49\textwidth]{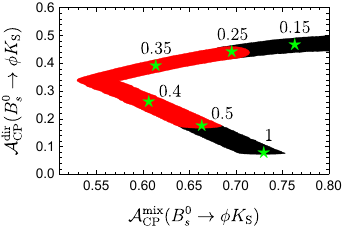}
\caption{SM correlations from Ref.~\cite{Fleischer:2026phiK} between the direct CP asymmetry and the hadronic phase shift in $\Bd\to\phi\KS$ (left), and between the direct and mixing-induced CP asymmetries in $\Bs\to\phi\KS$ (right). Red (black) regions correspond to momenta inside (outside) the range $1/4<k^2/m_b^2<1/2$ and stars mark selected values of $k^2/m_b^2$.}
\label{fig:sm-correlations}
\end{figure}

Despite their similar quark-level structure, $\Bd\to\phi\KS$ and $\Bs\to\phi\KS$ are not related by a clean $U$-spin correspondence. The dominant QCD penguin contribution involves a $B\to K$ transition in the former and a $B_s\to\phi$ transition in the latter. Comparable penguin parameters are nevertheless expected. Taking $\widetilde b=b_{\rm incl}$, $\widetilde\theta=\theta$ and $\phi_s=-(2.01\pm0.12)^\circ$~\cite{Barel:2023penguins} gives the SM benchmarks
\begin{equation}
\label{eq:bs-results}
\Adir=0.32\pm0.10\ ,\qquad \Amix=0.64\pm0.04\ ,\qquad \ADG=0.69\pm0.07\ .
\end{equation}
Figure~\ref{fig:sm-correlations} shows the correlations between $\Adir$--$\Delta\phi_d^{\phi\KS}$ (left) and $\Amix$--$\Adir$ (right). In contrast to the CP asymmetries which require tagging information, the decay-width-difference observable $\ADG$ can be accessed through the untagged decay-time distribution.

For the branching ratio, non-factorizable corrections to the dominant penguin amplitude are parametrized by $a_{\NF}^s$. Assuming $a_{\NF}^s=a_{\NF}^0$, with $a_{\NF}^0$ extracted from the measured $\Bd\to\phi\KS$ rate within the SM, gives
\begin{equation}
\label{eq:bs-br}
\BR_{\rm theo}=(8.6\pm3.1)\times10^{-8}\ ,\qquad 
\BR_{\rm exp}=\frac{1+\ADG y_s}{1-y_s^2}\,\BR_{\rm theo}=(9.0\pm3.3)\times10^{-8}\ ,
\end{equation}
where the conversion to the time-integrated branching ratio $\BR_{\rm exp}$ accounts for the sizeable decay width difference $y_s\equiv\Delta\Gamma_s/(2\Gamma_s)=0.062\pm0.004$~\cite{DeBruyn:2012BR}.
No measurements of this channel were available at the time of the analysis in Ref.~\cite{Fleischer:2026phiK}.
A branching-ratio measurement would probe the amplified penguin contributions, while CP-asymmetry measurements would also test the factorization-based estimates. Experimental studies of $\Bs\to\phi\KS$ are therefore strongly encouraged.

\clearpage
\section{Isospin observables from neutral and charged modes}
\label{sec:isospin}

The dominant $\Bd\to\phi K^0$ and $B^+\to\phi K^+$ topologies differ only through their spectator quarks. Decomposing the effective Hamiltonian into $I=0$ and $I=1$ components gives amplitudes of the form $A_{(0)}\pm A_{(1)}$, where the upper (lower) sign refers to the neutral (charged) mode~\cite{Fleischer:2001isospin}. The corresponding hadronic parameters $b_\pm$ and $\theta_\pm$ generalise those in Eq.~\eqref{eq:amplitude-sm}, with their subscripts matching the sign in $A_{(0)}\pm A_{(1)}$. Electroweak and dynamical suppression imply $|P_{(1)}^{(ct)\prime}/P_{(0)}^{(ct)\prime}|=\mathcal O(\bar\lambda^2)$, with $\bar\lambda\simeq0.2$ denoting a generic expansion parameter. Within this estimate, $b_+-b_-=\mathcal O(\bar\lambda^2)$.

The experimentally accessible combination
\begin{equation}
\label{eq:U}
\Uobs=\frac{\BR(B^+\to\phi K^+)}{2\BR(\Bd\to\phi\KS)}
\frac{\tau_{\Bd}}{\tau_{B^+}}
\frac{m_{\Bd}^3\Phi_0^3}{m_{B^+}^3\Phi_+^3}
\end{equation}
accounts for lifetime and phase-space differences. Here $\tau_{\Bd}$ and $\tau_{B^+}$ are the meson lifetimes and $\Phi_i=\Phi(m_{K_i}/m_{B_i},m_\phi/m_{B_i})$, with $\Phi(x,y)=\sqrt{[1-(x+y)^2][1-(x-y)^2]}$.
The factor of two converts the $\KS$ rate into the $K^0$ rate, neglecting CP violation in the neutral-kaon system. Isospin-breaking effects in the form factors and non-factorizable normalizations are collected in
\begin{equation}
\label{eq:WZ}
1+\Wobs=\left[\frac{F_1^{B^0K^0}(m_\phi^2)}{F_1^{B^+K^+}(m_\phi^2)}\frac{a_{\NF}^0}{a_{\NF}^+}\right]^2,\qquad \Zobs=\frac{1-\Uobs(1+\Wobs)}{1+\Uobs(1+\Wobs)}\ .
\end{equation}
Allowing corrections at the one-per-cent level in each amplitude ratio gives $\Wobs=\mathcal O(4\%)$. After accounting for these effects, $\Zobs$ probes the difference between the hadronic penguin parameters. The short-distance estimate is particularly small, $\Zobs_{\SM}=\mathcal O(\eps\bar\lambda^2)\sim0.1\%$. We allow a conservative $\mathcal O(1\%)$ to accommodate possible rescattering enhancements~\cite{Fleischer:2026phiK}.

Using the experimental inputs in Ref.~\cite{Fleischer:2026phiK} give
\begin{equation}
\label{eq:Zresult}
\Uobs=1.07\pm0.13,\qquad \Zobs=-0.03\pm0.07\ ,
\end{equation}
where the extraction of $\Zobs$ also uses the above estimate for $\Wobs$. The result for $\Zobs$ is consistent with the SM. Its interpretation as a probe of isospin-dependent NP therefore requires improved experimental precision as well as control over $\Wobs$.

Two further isospin observables can be constructed from the direct CP asymmetries \cite{Fleischer:2001isospin}:
\begin{equation}
\label{eq:SD}
\begin{aligned}
\Sobs&=\tfrac12\left[\Adir(\Bd\to\phi\KS)+\Adir(B^+\to\phi K^+)\right]\ ,\\
\Dobs&=\tfrac12\left[\Adir(\Bd\to\phi\KS)-\Adir(B^+\to\phi K^+)\right]\ .
\end{aligned}
\end{equation}
Using the current experimental inputs, we find
\begin{equation}
\label{eq:SDexp}
\Sobs_{\exp}=-0.04\pm0.06\ ,\qquad \Dobs_{\exp}=-0.05\pm0.06\ .
\end{equation}
The corresponding SM estimates are
\begin{equation}
\label{eq:SDsm}
\Sobs_{\SM}=-0.014\pm0.005\ ,\qquad \Dobs_{\SM}=0\pm0.003\ .
\end{equation}
The experimental results are compatible with these predictions, but their uncertainties still leave substantial room for NP contributions with different isospin structures.

\clearpage
\section{Probing isospin-dependent New Physics}
\label{sec:np}

To illustrate the sensitivity to NP, we retain one effective contribution in each isospin sector. The amplitudes can then be written as~\cite{Fleischer:2001isospin, Fleischer:2026phiK}
\begin{equation}
\label{eq:np-amplitude}
A(B^{0,+}\to\phi K^{0,+})=\frac{\mathcal A'_{P,\pm}}{\sqrt{\eps}}\left[1+\eps b_\pm e^{i\theta_\pm}e^{i\gamma}+v_0e^{i\Delta_0}e^{i\Phi_0}\pm v_1e^{i\Delta_1}e^{i\Phi_1}\right]\ ,
\end{equation}
where $v_I$ is the NP magnitude relative to the leading SM amplitude, $\Phi_I$ is a CP-violating weak phase, and $\Delta_I$ is a CP-conserving strong phase. To first order in the NP amplitudes, neglecting their interference with the doubly Cabibbo-suppressed SM term yields
\begin{equation}
\label{eq:np-observables}
\Sobs_{\NP}\simeq-2v_0\sin\Delta_0\sin\Phi_0\ ,\qquad
\Dobs_{\NP}\simeq-2v_1\sin\Delta_1\sin\Phi_1\ ,\qquad
\Zobs_{\NP}\simeq2v_1\cos\Delta_1\cos\Phi_1\ .
\end{equation}
Thus $\Sobs$ probes the $I=0$ sector, whereas $\Dobs$ and $\Zobs$ probe $I=1$ contributions. The suppressed SM term is retained in the numerical analysis.

\begin{figure}[htb]
\centering
\begin{minipage}{.48\textwidth}\centering
\small $I=0$, $\Delta_0=174^\circ$\\[-3pt]
\includegraphics[clip,width=\linewidth]{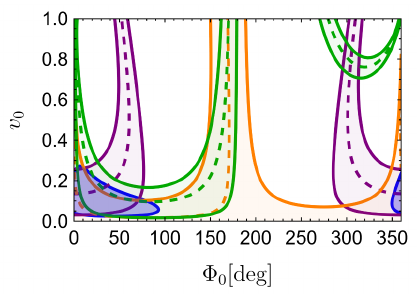}
\end{minipage}\hfill
\begin{minipage}{.48\textwidth}\centering
\small $I=1$, $\Delta_1=187^\circ$\\[-3pt]
\includegraphics[clip,width=\linewidth]{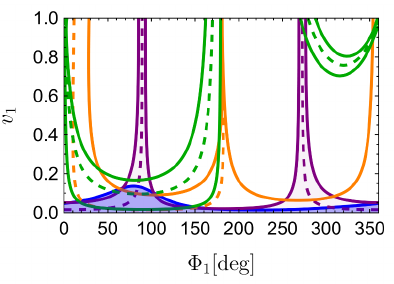}
\end{minipage}
\caption{NP constraints from Ref.~\cite{Fleischer:2026phiK}, with the strong phases fixed to the indicated best-fit values. Orange (green) regions follow from charged-mode direct (neutral-mode mixing-induced) CP violation. Purple regions use $\Delta\BR$ with exclusive CKM inputs for $I=0$, and $\Zobs$ for $I=1$. Blue regions are the $1\sigma$ best-fit regions.}
\label{fig:np}
\end{figure}

Figure~\ref{fig:np} shows each isospin sector separately. Both scenarios use direct and mixing-induced CP violation. The $I=0$ fit additionally uses $\Delta\BR\equiv\BR_{\rm exp}/\BR_{\SM}^{\rm fact}$ for $\Bd\to\phi\KS$, while for $I=1$ this is replaced by $\Zobs$, which carries a smaller theoretical uncertainty. Contributions up to $v_0\lesssim0.3$ and $v_1\lesssim0.2$ remain allowed in these benchmark analyses. 

\paragraph{Conclusions and outlook.}
Penguin-dominated $B\to\phi K$ decays offer not only sensitive tests of the SM, but also probes of the isospin structure of possible NP contributions. Current data are consistent with the SM benchmarks, yet still allow sizeable NP amplitudes. Observation and CP studies of $\Bs\to\phi\KS$ would provide new experimental handles on the hadronic penguin dynamics, while improved determinations of $\Sobs$, $\Dobs$ and $\Zobs$ would help disentangle possible $I=0$ and $I=1$ NP contributions. This complementarity makes the $B\to\phi K$ system a compelling target for precision measurements at LHCb and Belle~II, with the benchmarks set here guiding future NP searches.

\paragraph{Acknowledgments.}
This research is supported by NWO. The work of K.K.V. is supported in part by the NWO Vidi project \emph{Solving Beautiful Puzzles} (VI.Vidi.223.083).

\clearpage


\begin{thebibliography}{99}

\bibitem{Fleischer:1996penguins}
R.~Fleischer,
``CP violation and the role of electroweak penguins in nonleptonic $B$ decays,''
Int.\ J.\ Mod.\ Phys.\ A \textbf{12} (1997) 2459
[\href{https://arxiv.org/abs/hep-ph/9612446}{hep-ph/9612446}].

\bibitem{Grossman:1997phiK}
Y.~Grossman, G.~Isidori and M.P.~Worah,
``CP asymmetry in $B_d\to\phi K_S$: Standard model pollution,''
Phys.\ Rev.\ D \textbf{58} (1998) 057504
[\href{https://arxiv.org/abs/hep-ph/9708305}{hep-ph/9708305}].

\bibitem{Grossman:2003SU3}
Y.~Grossman, Z.~Ligeti, Y.~Nir and H.~Quinn,
``SU(3) relations and the CP asymmetries in $B$ decays to
$\eta'K_S$, $\phi K_S$ and $K^+K^-K_{(S)}$,''
Phys.\ Rev.\ D \textbf{68} (2003) 015004
[\href{https://arxiv.org/abs/hep-ph/0303171}{hep-ph/0303171}].

\bibitem{Fleischer:2001isospin}
R.~Fleischer and T.~Mannel,
``Exploring new physics in the $B\to\phi K$ system,''
Phys.\ Lett.\ B \textbf{511} (2001) 240
[\href{https://arxiv.org/abs/hep-ph/0103121}{hep-ph/0103121}].

\bibitem{Biswas:2024phiK}
A.~Biswas, S.~Descotes-Genon, J.~Matias and G.~Tetlalmatzi-Xolocotzi,
``Optimised observables and new physics prospects in the penguin-mediated decays
$B_{d(s)}\to K^{(*)0}\phi$,''
JHEP \textbf{08} (2024) 030
[\href{https://arxiv.org/abs/2404.01186}{arXiv:2404.01186}].


\bibitem{Fleischer:2026phiK}
R.~Fleischer, J.~Groot and K.~K.~Vos,
\emph{CP violation in $B_{(s)}\to\phi K$ decays: Standard Model benchmarks and isospin-breaking New Physics},
\href{https://doi.org/10.1007/JHEP09(2026)165}{JHEP \textbf{09} (2026) 165}
[\href{https://arxiv.org/abs/2603.13139}{arXiv:2603.13139}].

\bibitem{Fleischer:1994}
R.~Fleischer,
\emph{Electroweak Penguin effects beyond leading logarithms in the $B$ meson decays $B\to K^-\phi$ and $B^-\to\pi^-\bar K^0$},
Z.~Phys.~C \textbf{62} (1994) 81.

\bibitem{LHCb:2021gamma}
LHCb collaboration,
\emph{Simultaneous determination of CKM angle $\gamma$ and charm mixing parameters},
JHEP \textbf{12} (2021) 141
[\href{https://arxiv.org/abs/2110.02350}{arXiv:2110.02350}].

\bibitem{Barel:2023penguins}
M.~Z.~Barel, K.~De~Bruyn, R.~Fleischer and E.~Malami,
\emph{Penguin Effects in $B_d^0\to J/\psi K_{\rm S}^0$ and $B_s^0\to J/\psi\phi$},
PoS \textbf{CKM2021} (2023) 111
[\href{https://arxiv.org/abs/2203.14652}{arXiv:2203.14652}].

\bibitem{DeBruyn:2012BR}
K.~De~Bruyn et al.,
\emph{Branching Ratio Measurements of $B_s$ Decays},
Phys.~Rev.~D \textbf{86} (2012) 014027
[\href{https://arxiv.org/abs/1204.1735}{arXiv:1204.1735}].

\end{thebibliography}
\end{document}